\documentclass[11pt,a4paper]{article} 
\usepackage{jheppub,hyperref,mdwlist}
\usepackage{jheppub,hyperref,mdwlist}
\usepackage{amsmath}
\usepackage{graphicx}
\usepackage{subfigure}
\usepackage{amssymb}
\usepackage{xcolor}
\usepackage{multirow}
\usepackage{cancel}
\usepackage{color}
\usepackage{mathtools}
\usepackage{listings}
\usepackage{xcolor}
\usepackage{float}
\usepackage{slashed}
\usepackage{bm}
\usepackage{amsmath}
\usepackage{wrapfig}

\newcommand{\be}{\begin{equation}}
\newcommand{\ee}{\end{equation}}
\newcommand{\beq}{\begin{equation}}
\newcommand{\eeq}{\end{equation}}
\newcommand{\bea}{\begin{eqnarray}}
\newcommand{\eea}{\end{eqnarray}}
\definecolor{airforceblue}{rgb}{0.36, 0.54, 0.66}
\definecolor{steelblue}{rgb}{0.27, 0.51, 0.71}
\definecolor{amber}{rgb}{1.0, 0.49, 0.0}

\title{\boldmath Non-thermal dark  matter production and leptogenesis in a $L_{\mu}-L_{\tau}$ model}

\author{XinXin Qi,}
\author{Hao Sun}
\affiliation{Institute of Theoretical Physics, School of Physics, Dalian University of Technology, No.2 Linggong Road, Dalian, Liaoning, 116024, P.R.China }
\emailAdd{qxx@mail.dlut.edu.cn}
\emailAdd{haosun@dlut.edu.cn}

\abstract{
  We discuss the possibility of light scalar dark matter generated by right-handed neutrino in a $L_{\mu}-L_{\tau}$ model, in which the dark matter $\phi_{dm}$ carries $U(1)_{L_{\mu}-L_{\tau}}$ charge but it is a singlet in the Standard Model. We discuss the case that dark matter production mainly comes from  scattering associated with a pair of right-handed neutrinos non-thermally while other related processes are highly suppressed. A feasible parameter space is considered and we found the correct dark matter relic density can be obtained without influencing the result of leptogenesis result. The heavier right-handed neutrino will induce colder dark matter production and the allowed dark matter mass region is $[\rm 10^{-5}\ GeV,0.1\ GeV]$.
}

\begin{document}
\maketitle
\flushbottom
\section{Introduction}
\label{sec:intro}

The Standard Model (SM) is regarded as one of the most successful theories in physics, which can describe strong and electroweak interactions in an elegant way. However, there are still other open questions the SM can not explain. Neutrino oscillation experiments \cite{Super-Kamiokande:1998kpq,SNO:2002tuh} indicate that neutrinos are not massless, and the mass of neutrinos are at the sub-eV level, which is contrary to the SM. 
According to astronomical exploration, our universe is not only composed of baryon matter, but also composed of dark matter (DM) and dark energy. The combined analysis shows that the dark matter density and baryon density are \cite{Planck:2018vyg}:
   \begin{eqnarray}
       \Omega h^2=0.12 \pm 0.001  \ {\text{and}} \ \Omega_b h^2=0.0224 \pm 0.0001 ,
   \end{eqnarray}
respectively.
In addition, the number of baryons in the universe is not equal to the number of anti-baryons. The baryon asymmetry can be expressed by \cite{Planck:2018vyg}:
   \begin{eqnarray}
       \eta \equiv \frac{n_B-n_{\bar{B}}}{n_{\gamma}}|_0=(6.12 \pm 0.03)\times 10^{-10}  
         \quad 
\end{eqnarray}
or  
\begin{eqnarray}
     Y_B \equiv \frac{n_B-n_{\bar{B}}}{s}|_0 \approx 8.7 \times 10^{-11}
\end{eqnarray}
where $n_B$, $n_{\bar{B}}$, $n_{\gamma}$ and $s$ are the number densities of baryons, anti-baryons, photons and entropy, respectively. The subscript 0 means at present time. 

The evidence of neutrino mass, dark matter and baryon asymmetry in the universe provide us with new hints beyond SM. In particular, by introducing right-handed neutrinos, the problems of neutrino mass and baryon asymmetry can be solved 
simultaneously with leptogenesis \cite{Fukugita:1986hr}, where the light neutrino mass can be generated via the type-I seesaw  mechanism \cite{Minkowski:1977sc,Mohapatra:1979ia,Gell-Mann:1979vob,Schechter:1980gr,Glashow:1979nm}, and the decays of right-handed neutrinos in the early universe can produce the baryon asymmetry naturally.
Concretely speaking, a lepton asymmetry can be generated through the CP-violating and out-of-equilibrium decay of right-handed neutrinos in the early universe during leptogenesis, and this asymmetry can be converted into baryon asymmetry via $\rm (B+L)$-violating interactions during the electroweak symmetry breaking (EWSB)  phase transition. 

Attempts at models relating dark matter to baryon asymmetry have been discussed in Refs.\cite{Nussinov:1985xr,Kaplan:1991ah,Farrar:2005zd,Hooper:2004dc,Kitano:2004sv,Agashe:2004bm,Kitano:2008tk,Nardi:2008ix,Kaplan:2009ag,Cohen:2010kn,Hall:2010jx,Feldstein:2010xe,Shelton:2010ta,Davoudiasl:2010am,Haba:2010bm,Gu:2010ft,Blennow:2010qp,McDonald:2010toz,Cheung:2010gj,Cheung:2010gk,Dutta:2010va,Zurek:2013wia,Hall:2019rld,Hall:2021zsk}, and among these models a two sector leptogenesis scenario is introduced to produce comparable baryon and dark matter number densities based on the fact of $\Omega h^2 \approx 5\Omega_b h^2$. Such a scenario connects dark matter and leptogenesis together, in which the heavy right-handed neutrinos couple to the dark sector as well as the SM leptons in the visible sector, which results in an asymmetry in both dark and visible sectors simultaneously. Typically, as discussed in Refs.\cite{Falkowski:2017uya,Falkowski:2011xh}, the two-sector leptogenesis scenario can also  relate light dark matter with  leptogenesis and predicts light dark matter mass ranging from keV to GeV without matching asymmetries between visible and dark sector. Imitating a similar idea, we assume SM particles and dark matter are related via Yukawa interactions to the same  right-handed neutrinos, and the  right-handed neutrinos provide a common origin of the light neutrino mass (type-I seesaw mechanism), leptogenesis (decay of heavy right-handed neutrinos related with CP violation) and dark matter (heavy neutrinos generate dark matter via Yukawa interactions). 
   
In principle, the shared production mechanism to generate baryon asymmetry and the dark matter relic density in our universe often demands adding new particles to the SM. For example, the additional Yukawa  terms can connect right-handed neutrinos and dark matter with an extra scalar field which is SM singlet. Such singlet scalar field can be unified in concrete models and introduce new phenomenology at the LHC experiments. One of the most attractive models including such new scalar fields is  $L_{\mu}-L_{\tau}$ extension of the SM \cite{Foot:1994vd,Altmannshofer:2016brv,Foldenauer:2016rpi,Baek:2001kca,He:1990pn, Heeck:2010pg,Chen:2017gvf,Asai:2017ryy,Qi:2021rhh,Araki:2019rmw},
which involves a new gauge boson $Z_p$, and a singlet scalar with $L_{\mu}-L_{\tau}$ charge. The new light gauge boson $Z_p$ can explain the $(g-2)_{\mu}$ anomaly \cite{Muong-2:2021ojo,Muong-2:2021ovs,Muong-2:2021vma} naturally considering the one-loop contribution of $Z_p$ to muon's magnetic dipole momentum  \cite{Heeck:2010pg}.
Discussion about dark matter in the  $L_{\mu}-L_{\tau}$ model can be found in Refs.\cite{Foldenauer:2018zrz,Altmannshofer:2016jzy,Biswas:2016yjr,Das:2021zea,Biswas:2016yan,Deka:2022ltk,Costa:2022oaa,Okada:2019sbb} while leptogenesis in the $L_{\mu}-L_{\tau}$ model can be found in Refs.\cite{Chun:2007vh,Adhikary:2006rf,Asai:2020qax}.
   
In this work, we consider dark matter carries $U(1)_{L_{\mu}-L_{\tau}}$ charge and is a singlet in the SM. We discuss the possibility of non-thermal dark matter production generated by right-handed neutrinos at the early universe, which contributes to the observed dark matter relic density. What's more, we focus on the relic density constraint on the  parameter space of scatterings within the non-thermal mechanism. 
For the above issues, we have the following comments: Firstly, the dark matter $\phi_{dm}$ has several production channels: gauge portal, Higgs portal and Yukawa portal. For the Higgs portal, one can fine-tune the scalar-DM couplings to be relatively small so that the contribution of the Higgs portal channels to dark matter production can be negligible. For the gauge portal, dark matter carries $U(1)_{L_{\mu}-L_{\tau}}$ charge and 
a high scale breaking $L_{\mu}-L_{\tau}$ symmetry is  considered, which leads to a tiny $U(1)_{{L_\mu}-{L_\tau}}$ charge of dark matter, and the gauge portal contribution will be highly suppressed due to the choice of the small charge. Therefore, Yukawa portal channels can play an important role in determining dark matter abundance. Note that right-handed neutrinos can not only generate the baryon asymmetry via the leptogenesis mechanism  but also dark matter simultaneously in the model. On the other hand, there can be no dark matter initially and dark matter particles can be produced through the decay or annihilation of heavier particles (right-handed neutrinos in this work) that were present in the early Universe, and the non-thermal production  can contribute to the dark matter abundance. Unlike most two-sector leptogenesis scenarios, where sterile neutrinos as dark matter are coupled to the scalar field in the dark sector and right-handed neutrinos through the Yukawa term and the dark matter abundance is derived  through the decays of right-handed neutrinos during the leptogenesis process \cite{Falkowski:2017uya}. In this work, we consider a scheme in which a singlet scalar is the dark matter coupling directly to the right-handed neutrinos. Furthermore, the abundance of dark matter is mainly given by the scattering process during leptogenesis, 
which leads to  dependence of the dark matter abundance with the singlet Yukawa interactions instead of the branching ratio\cite{Bandyopadhyay:2017bgh}. 
%



The article is arranged as followed, in section \ref{sec:2}, we give the scalar dark matter model based on the $L_{\mu} -L_{\tau}$ framework. In section \ref{sec:3}, we briefly discuss the $(g-2)_{\mu}$ anomaly in the model. In section \ref{sec:4}, we discuss the dark matter phenomenology as well as leptogenesis in the model, and finally we summarize in the last section of the paper.

\section{Model description}
\label{sec:2}
In the context of the minimal $L_{\mu}-L_{\tau}$ model with only one extra singlet scalar \cite{Adhikary:2006rf,Asai:2020qax}, the $\mu\mu$ and $\tau\tau$ entries of the right-handed neutrino mass matrix are equal to zero, which leads to a so-called two-zero minor structure \cite {Asai:2017ryy} in the light neutrino mass matrix. According to the global fit data  \cite{ParticleDataGroup:2020ssz,Esteban:2018azc,deSalas:2020pgw} 
and cosmology constraints \cite{Planck:2018vyg} on the sum of light neutrino masses, it is  difficult to satisfy the data in such a minimal model as discussed in Refs.\cite{Asai:2017ryy,Asai:2020qax}. In this work, we consider one  possible extension of the minimal $L_{\mu}-L_{\tau}$ model \cite{Borah:2021mri} including scalar dark matter $\phi_{dm}$ as followed in Table \ref{tab2}.
\noindent
\begin{table}[!h]
	\begin{center}
		\footnotesize
		\begin{tabular}{||@{\hspace{0cm}}c@{\hspace{0cm}}|@{\hspace{0cm}}c@{\hspace{0cm}}|@{\hspace{0cm}}c@{\hspace{0cm}}|@{\hspace{0cm}}c@{\hspace{0cm}}||}
			\hline
			\hline
			\begin{tabular}{c}
				{\bf Gauge group}\\
			    { }\\
				\hline
				 
				$\tiny{SU(2)_{L}}$\\ 
				\hline
				$U(1)_{Y}$\\ 
				\hline
				$U(1)_{L_\mu-L_\tau}$\\ 
			\end{tabular}
			&
			&
			\begin{tabular}{c|c|c}
				\multicolumn{3}{c}{\bf Fermion Fields}\\
				\hline
			$N_e$& $N_{\mu}$& $N_{\tau}$ \\
				\hline
				$1$&$1$&$1$\\
				\hline
				$0$&$0$&$0$\\
				\hline
				$0$&$q$&$-q$\\
			\end{tabular}
			&
			\begin{tabular}{c|c|c|c}
				\multicolumn{3}{c}{\bf Scalar Field}\\
				\hline
				 ~~$H_2$~~ & ~~$H_3$~~ & ~~$\phi_{1,2}$~~ & ~~$\phi_{dm}$ ~~ \\
				\hline
				$2$&$2$&$1$&0\\
				\hline
				$1/2$&$1/2$&$0$&$1$\\
				\hline
				$1-q$&$-1+q$&$q,2q$&$2q$\\
			\end{tabular}\\
			\hline
			\hline
		\end{tabular}
		\caption{
New Particles and their respective gauge charges in the non-minimal model.}
		\label{tab2}
	\end{center}    
\end{table}

We introduce three right-hand neutrinos $N_e$, $N_{\mu}$ and $N_{\tau}$ to the SM, and the corresponding $U(1)_{L_{\mu}-L_{\tau}}$ charge are 0, $q$ and $-q$ respectively, where $q$ is a certain number. We also introduce two doublet scalar fields $H_2$ and $H_3$ with $U(1)_{L_{\mu}-L_{\tau}}$ charge $1-q$ and $q-1$ as well as two singlet scalars with $U(1)_{L_{\mu}-L_{\tau}}$ charge $q$ and $2q$, the singlet scalar dark matter $\phi_{dm}$ charge is $2q$.

Lagrangian related with neutrino mass as well as dark matter can be written as:

\begin{align}
\mathcal{L} & \supseteq  -y_{sm}\phi_{dm}N_{\mu}N_{\mu}- y_{st}\phi_{dm}^{\dagger}N_{\tau}N_{\tau}- \frac{M_{\mu \tau}}{2} N_{\mu} N_{\tau}  - \frac{M_{ee}}{2} N_e N_e -y_{e\mu} \phi^{\dagger}_1 N_e N_\mu -y_{e\tau} \phi_1 N_e N_\tau \notag \\
 & -y_{\mu} \phi^{\dagger}_2 N_\mu N_\mu -y_{De} \overline{L}_e \widetilde{H} N_e-y_{D\mu} \overline{L}_\mu \widetilde{H}_2 N_\mu-y_{D \tau} \overline{L}_\tau \widetilde{H}_3 N_\tau-y_{\tau} \phi_2 N_\tau N_\tau - y_{le} \overline{L}_e H e_R \notag \\
 & + y_{l\mu} \overline{L}_\mu H \mu_R + y_{l\tau} \overline{L}_\tau H \tau_R+ {\rm h.c.},
\label{yuklag}
\end{align}

where $\widetilde{H}=i\sigma_2H$, $\widetilde{H}_2=i\sigma_2H_2$ and $\widetilde{H}_3=i\sigma_2H_3$.  
There are three Higgs doublets and two SM singlets in such model, and the scalar doublet $H$ plays the role of SM Higgs. 
We focus on the interplay between the dark matter and leptogenesis while the couplings of dark matter with other scalar fields such as $|\phi_{dm}|^2|H|^2$ are negligible. On the other hand, $\phi_{dm}$  can have quadratic and trilinear terms with other singlet scalars such as $\phi_1^2\phi_{dm}^\dagger$,$\phi_{dm}^\dagger\phi_2$, which should be fine-tunning to satisfy dark matter lifetime bound. For simplicity, we set these couplings to be zero.
The doublets will be massless at the scale of leptogenesis because of the unbroken electroweak symmetry, while the singlet scalars are massive since the scale of $U(1)_{L_{\mu}-L_{\tau}}$ symmetry breaking is above the leptogenesis scale in the model. 
Although the singlet scalars play no role in leptogenesis, their masses are constrained by LHC experiments due to the mixings with SM like Higgs boson \cite{Robens:2015gla,Chalons:2016jeu}. 
The most stringent constraint on the mixing angle $\theta$ arises from the $W$ boson  mass correction \cite{Lopez-Val:2014jva} at NLO with 
$ 0.2 \lesssim sin\theta \lesssim 0.3$ in the case of 250 GeV $\lesssim m_{\phi_i} \lesssim $ 850 GeV, where $m_{\phi_i}$ is the mass of the singlet scalar $\phi_i$.
For $m_{\phi_i}< 250$ GeV, the LHC and LEP direct search \cite{CMS:2015hra,Strassler:2006ri} and measured Higgs signal strength \cite{Strassler:2006ri} yield $sin\theta \lesssim 0.25$. 
For $m_{\phi_i}> 850$ GeV, we have $sin\theta \lesssim 0.2$ from the requirement of perturbativity and unitarity. 
In summary, the bounds are mild for $m_{\phi_i}< 1$ TeV. Since the singlet scalars make no difference in leptogenesis, the mixing angle can remain as small as required by tuning their couplings with SM like Higgs.  On the other hand, the physical scalars from the Higgs doublets are also constrained by experiments. Similar with the singlet scalar case, we can satisfy these bounds by tuning the related parameters and without changing the results about leptogenesis. In this work, we don't elaborate on them further. 
 
For simplicity, we assume the vevs of these scalar fields are given by,
\begin{eqnarray}
&& \langle H\rangle=v_0,\ \langle H_2\rangle =\langle H_3 \rangle=v_h, \\
&&  \langle \phi_1\rangle =\langle \phi_2\rangle =v_b
\end{eqnarray}
and we have,
\begin{eqnarray}
v_0^2+2v_h^2=246^2,\ M_{Z_p}=\sqrt{5}qg_{p}v_b 
\label{eq1}
\end{eqnarray}

After spontaneous symmetry breaking, the right-handed neutrino mass matrix and Dirac neutrino mass matrix are given by,
\begin{eqnarray}
  M_R= \left(
  \begin{array}{ccc}
 
               M_{ee}      &  y_{e\mu} \frac{v_b}{\sqrt{2}}
    & y_{e\tau}  \frac{v_b}{\sqrt{2}} \\
               y_{e\mu}  \frac{v_b}{\sqrt{2}}      &  \sqrt{2} y_{\mu}
 v_b     & \frac{M_{\mu \tau}}{2}  \\
               y_{e\tau}  \frac{v_b}{\sqrt{2}}     &  \frac{M_{\mu \tau}}{2}    &
               \sqrt{2} y_{\tau}  v_b  .
    \end{array}
     \right)
\end{eqnarray}
and
\begin{eqnarray}
      M_D = \left(
      \begin{array}{ccc}
    y_{De} \frac{v_0}{\sqrt{2}}       &  0    & 0  \\
               0     & y_{D\mu} \frac{v_h}{\sqrt{2}}    & 0  \\
               0     &  0    & y_{D \tau} \frac{v_h}{\sqrt{2}}  
    \end{array} 
       \right)
 \end{eqnarray}
respectively. The light neutrino mass matrix $M_{\nu}$ can be generated by the Type-I seesaw mechanism with 
\begin{eqnarray}
M_{\nu}=-M_D M_R^{-1} M_D^{T}= U^*\hat{M}_{\nu}U^\dagger,
\end{eqnarray}
where $\hat{M}_{\nu}$= $diag\{m_1,m_2,m_3\}$  with $m_1,m_2$ and $m_3$ being the light neutrino mass, and $U$ is the Pontecorvo-Maki-Nakagawa-Sakata (PMNS) matrix \cite{Pontecorvo:1957cp,Maki:1962mu}.
In order to simplify the following discussion,  we take the following benchmark values of the PMNS matrix with \cite{Xing:2020ald},

\begin{align}
  U&=  
\left(
\begin{array}{ccc}
 0.82571 &  0.82571 & -0.124534-0.0846333 i \\
 -0.354387-0.0454376 i & 0.683844 -0.0299151 i & 0.635459 \\
 0.433147 -0.0541505 i & -0.484417-0.0356514 i &0.757311 \\
\end{array}
\right)
\end{align}

For the light neutrino mass, the two distinctive neutrino mass squared differences can be given by \cite{Xing:2020ald},
 \begin{eqnarray}
 &&  \Delta m_{21}^2=m_2^2-m_1^2=7.4\times 10^{-23} \rm \ GeV^2 \\
 && \Delta m_{31}^2=m_3^2-m_1^2=2.515\times 10^{-21} \rm \ GeV^2 
 \end{eqnarray}

\section{muon g-2}
\label{sec:3}

\begin{figure}[h]
 \centering
 \includegraphics[scale=0.65]{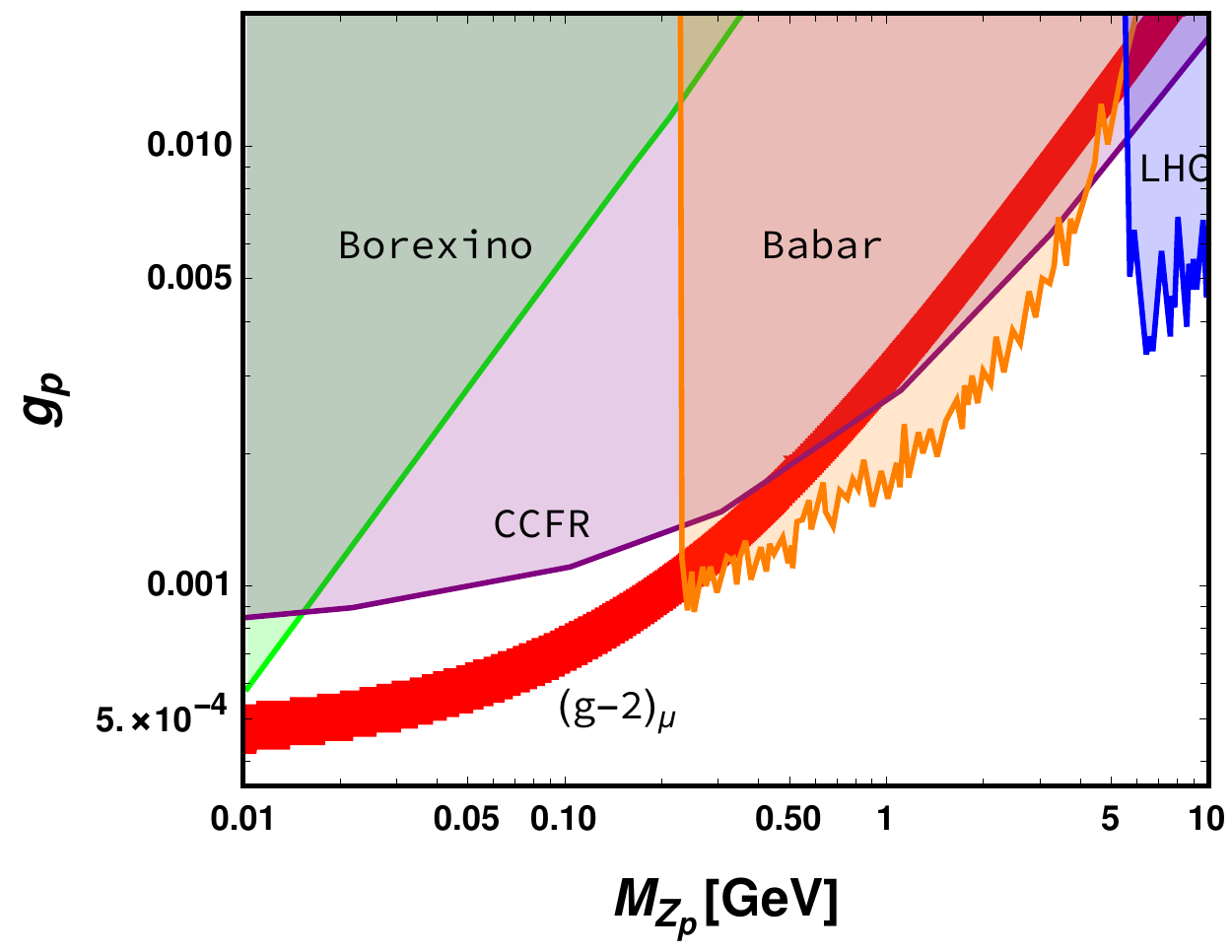}
 \caption{
  The allowed region that can explain the $(g-2)_{\mu}$ anomaly and the excluded region on $M_{Z_p}-g_p$ from different experiments:
  the green (orange, blue, purple) region is excluded by the Borexino (BABAR, LHC, CCFC).
  The red region is the parameter satisfying $(g-2)_{\mu}$ anomaly \cite{Qi:2021rhh}.
  }
\label{Fig:fig1}
\end{figure}

In the $L_{\mu}-L_{\tau}$ model, the $(g-2)_{\mu}$ anomaly can be solved naturally with the $Z_p$ one loop contribution. The analytical expression for $\Delta a_\mu$ is \cite{Baek:2008nz}
\begin{eqnarray}
 \Delta a_\mu = \frac{g_{p}^2 m_{\mu}^2}{8\pi^2} \int_{0}^{1} \frac{2\omega^2(1-\omega)}{\omega^2 m_{\mu}^2+(1-\omega)M_{Z_{p}}^2} d\omega
 \label{g-2}
\end{eqnarray}
where $m_{\mu}$ is the muon mass. We ignore the possible contribution of the additional Higgs doublets and focus on the gauge boson $Z_p$ as discussed in Ref.\ \cite{Borah:2021mri}.
Direct searches for new gauge bosons at experiments have already given stringent constraints on the parameter space of $(M_{Z_p},g_p)$.
At colliders, the $Z_p$ gauge boson can be produced via $e^+ e^- \to \mu^+ \mu^- Z_p$
with the subsequent decay of $Z_p\to\mu^+\mu^-$ \cite{Chun:2018ibr}. Such a search can be found in Babar \cite{BaBar:2016sci}
and give us the possible bound on $M_{Z_p}-g_p$ plane.
What's more, neutrino experiments give us new clues to constrain the parameter space \cite{Chakraborty:2021apc}.
The Borexino data related to the scattering of low energy solar neutrinos \cite{Borexino:2017rsf, Abdullah:2018ykz}
can provide the most stringent constraint on the low $M_{Z_p}$ and low $g_p$ region.
Another constraint is from CCFR collaboration \cite{Altmannshofer:2014pba}, obtained via the neutrino trident production where a muon neutrino scattered off of a nucleus producing a $\mu^+ \mu^-$ pair.
Such a process will be enhanced due to the existence of $Z_p$ compared with the SM case. The above experiments constrain the $g_{p}-M_{Z_p}$ parameter space with $M_{Z_p}\sim $ $(10,200)$ MeV and $g_p$ $\lesssim$ $10^{-4} \sim 10^{-3}$ level.
See details that are shown in Fig.\ref{Fig:fig1}. Experiment bounds on $g_p-M_{Z_p}$ indicate $q v_b$ in Eq.~\ref{eq1} is approximately equal to a hundred GeV.

\section{Dark matter from leptogenesis}
\label{sec:4}

\begin{figure}[h]
  \centering
  \includegraphics[height=5cm,width=5cm]{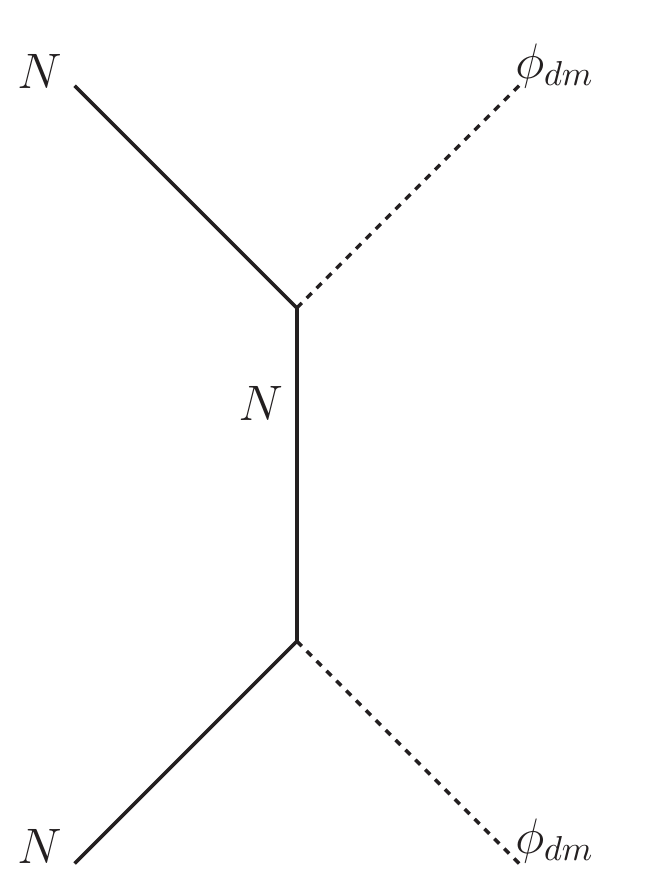}
  \caption{Processes related with the production of dark matter.}
  \label{Fig:fig2}
  \end{figure}

In this section, we consider dark matter relic density as well as leptogenesis in the model. According to Eq.~\ref{eq1}, $v_b$ and $q$ are related by a pair of $M_{Z_p}-g_p$, which are constrained by the $(g-2)_{\mu}$ anomaly. On the other hand, $v_b$ can determine right-handed neutrino mass scale while $q$ corresponds to  $U(1)_{L_{\mu}-L_{\tau}}$ charge of $N_{\mu}$ which is also related with dark matter. $M_{Z_p}$ and $g_p$ make little difference on our result in this work, therefore we choose $M_{Z_p}=0.2$ GeV and $g_p=0.001$ in the following discussion for simplicity, which satisfies $(g-2)_{\mu}$ anomaly as well as other collider experiment constraints.
The scalar-DM couplings are  fine-tuned to be negligible and the gauge-portal channels related to dark matter are highly suppressed due to the choice of the tiny charge $q$. Therefore, dark matter relic density mainly depends on the scatter processes of $N_{\tau(\mu)} N_{\tau(\mu)} \rightarrow \phi_{dm} \phi_{dm}$. For simplicity, we focus on the case of $N_{\tau}$, which also plays an important role in leptogenesis $^{\footnotemark[1]}$,
and the dominant process related to dark matter relic density is the t-channel interaction of 
$N_{\tau} N_{\tau} \rightarrow \phi_{dm} \phi_{dm}$ according to Fig.~\ref{Fig:fig2}. 
There exist couplings of dark matter with other particles in the model, which make the scalar field $\phi_{dm}$ unstable since $\phi_{dm}$ can decay into these particles. To make sure the scalar $\phi_{dm}$ is the dark matter, the lifetime of $\phi_{dm}$ should be longer than  $10^{27}s$, or equivalently $\Gamma_{\phi_{dm}} \leq 6.6 \times 10^{-52}$ GeV in terms of decay width \cite{Mohapatra:2020bze}. In the model, we have two possible decay processes related to dark matter. Firstly, the decay channels such as $\phi_{dm} \rightarrow N_{\tau} N_{\tau}$ are kinetically forbidden because $\phi_{dm}$ mass is much smaller than the heavy right-handed neutrinos. Then, we have couplings of dark matter with light neutrinos due to the heavy-light neutrino mixing, and the lifetime of dark matter is determined by $y_{st}$ and the mixing angle, where the mixing angle is highly suppressed by the right-handed neutrino mass.  

\footnotetext[1]{Contribution of $N_{\mu}$ to the leptogenesis can be negligible when $N_{\mu}$ mass is much heavier than $N_{\tau}$.}   

\subsection{Leptogenesis}  

In this part, we consider leptogenesis arising from the out-of-equilibrium decays of the heavy right-handed neutrinos in the model.  The generated (B-L) asymmetry can be converted into the baryon asymmetry via the sphaleron processes \cite{Manton:1983nd,Kuzmin:1985mm}, which are active between temperatures of $10^{12}$ GeV to 100 GeV in the early Universe. The  relevant Yukawa  matrix for leptogenesis in the model can be identified by \cite{Borah:2021mri}
\begin{eqnarray}
   y=\left(
  \begin{array}{ccc}
  y_{De}V_{11} & y_{De}V_{12} & y_{De}V_{13} \\
  y_{D \mu}V_{21} & y_{D\mu}V_{22} & y_{D\mu}V_{23} \\
  y_{D\tau}V_{31} &y_{D\tau}V_{32} & y_{D\tau} V_{33}\\
  \end{array} 
  \right)
\end{eqnarray}
where $V_{ij}$ are the elements of the matrix $V$ which diagonalizes $M_R$
\begin{equation}
{\rm diag}(M_{1},M_{2},M_{3})=V^{\dagger} M_{R}V^* .
\end{equation}
The light neutrino mass matrix poses no special structure and in this work the neutrino mass is fit with the current experiment data. In this part, we choose the following value of $M_R$ in the unit of GeV  to  fit the proper light neutrino mass.
\begin{align}
  M_R &= 10^{10} \times \left(
\begin{array}{ccc}
  -7.43984 + 0.134391 i& -0.963963+0.592564 i& -0.829834- 0.670913 i \\
 -0.963963+0.592564 i & -4.92313+0.118174 i & -0.518899+0.0157258 i \\
 -0.829834-0.670913 i & -0.518899+0.0157258 i& -4.74867-0.196062 i\\
\end{array}
\right)
\end{align}
where we have set $y_{De}=-0.02$, $y_{D\mu}=0.02$, $y_{D\tau}=-0.02$ and $v_0=200$ GeV. In addition, one can give the elements of $V$ numerically via the values of $M_R$, 
\begin{eqnarray}
V=\left(
\begin{array}{ccc}
 0.868993 & -0.339967-0.244011 i & -0.287073+0.248069 i \\
0.338398-0.142866 i& 0.0954554+0.161961 i &0.872793 \\
 0.270087+0.192304 i & 0.888559 & -0.271721-0.143009 i \\
\end{array}
\right)
\end{eqnarray}
so the matrix $y$ can be given explicitly by,
\begin{eqnarray}
 y =\left(
\begin{array}{ccc}
-2.457988 & 0.961572+0.690168 i & 0.811966-0.701645 i \\
 0.685471-0.289394 i & 0.193358+0.328074 i& 1.76796 \\
 -0.547097-0.389539 i & -1.7999 & 0.550407+0.289685 i \\
\end{array}
\right)
\end{eqnarray}

For convenience, we represent $N_{\tau}$ with $N$ in the following discussion. The reaction density for decay is given by,
\begin{eqnarray}
 \gamma_D(z)=\frac{m_{N}^3}{z\pi^2}K_1(z)\Gamma_{N}
\end{eqnarray}
where $K_1(x)$ is the first kind Bessel function, $z=m_N/T$ with T being temperature, and $\Gamma_N$ is the decay width of $N$.
   

To simplify the discussion, we consider dark matter relic density  generated by only  $N$ so that we fix $y_{sm}$ to be negligible. We also set the mass of  $N$ to be the lightest among the three heavy right-handed neutrinos. 
 We consider dark matter relic density mainly arising from the process of $NN\rightarrow \phi_{dm}\phi_{dm}$  in  Fig.~\ref{Fig:fig2} during leptogenesis, the reduced cross-section of $NN\rightarrow \phi_{dm}\phi_{dm}$ can be given by \cite{Dev:2017xry},
 \begin{align}
  \hat{\sigma}(x)&=\frac{y_{st}^4}{16\pi}(-2\beta_H(x)+\frac{4\beta_H(x)(r_H-4)^2}{x^2\beta_H(x)^2-(x-2r_H)^2}-
\frac{x^2-4(r_H-4)x+2(r_H-4)(3r_H+4)}{x(x-2r_H)} \notag \\
&\log\frac{(1-\beta_H(x))x-2r_H}{(1+\beta_H(x))x-2r_H}) \ \ \ ~~~
\label{eq25}
\end{align}
where 
\begin{eqnarray}
 x= \frac{s_s}{{m_N^2}}, ~r_H=\frac{m_{\phi_{dm}}^2}{m_N^2},~ \beta_H(x)=\sqrt{(1-4/x)(1-4r_H/x)}
\end{eqnarray}
and $s_s$ being the squared center-of-mass energy.
The reaction density for the process is given by \cite{Plumacher:1996kc},
\begin{eqnarray}
 \gamma(z)=\frac{m_N^4}{64\pi^4 z}\int_{4}^\infty \hat{\sigma}(\omega)\sqrt{\omega}K_1(z\sqrt{\omega}) .
\end{eqnarray}
  
The Boltzmann equations to describe abundance $Y_{N}$, dark matter abundance $Y_{dm}$ and lepton asymmetry $Y_{B-L}$ are given as followed in our model \cite{Davidson:2008bu},
\begin{eqnarray}
&&\frac{s_N H_N}{z^4}Y'_{N}=-(\frac{Y_{N}(z)}{Y_{Neq}(z)}-1)\gamma_D(z)-\frac{Y_{N}(z)^2}{Y_{Neq}(z)^2}2\gamma(z) \\
&&\frac{s_N H_N}{z^4}Y'_{dm}=\frac{Y_{N}(z)^2}{Y_{Neq}(z)^2}2\gamma(z) \label{b2}\\
&&\frac{s_N H_N}{z^4}Y'_{B-L}=-(\frac{1}{2}\frac{Y_{B-L}(z)}{Y_{leq}}+\epsilon(\frac{Y_{N}(z)}{Y_{Neq}(z)}-1))\gamma_D(z) -\Delta W Y_{B-L} 
\label{b3}
\end{eqnarray}
 where $\epsilon$ is the CP asymmetry parameter.
About quantities in the  above three equations, we have,
\begin{eqnarray}
   s_N=\frac{2\pi^2}{45}g_{*}m_{N}^3,\ H_N=\frac{1}{2\chi m_{pl}}m_{N}^2
\end{eqnarray}
where $m_{pl}$ is the Planck mass with $m_{pl}=1.22 \times 10^{19}$ GeV, $g_{*}$ is the effective degrees of freedom with $g_{*}=106.75$, and $\chi$ is defined by,
\begin{eqnarray}
\chi=\frac{1}{4\pi}\sqrt{\frac{45}{\pi g_*}} ,
\end{eqnarray}
$Y_{Neq}$ describes the abundance of $N$ at thermally equilibrium \cite{Davidson:2008bu},
\begin{eqnarray}
 Y_{Neq}(z)=\frac{45z^2}{2\pi^4g_{*}}K_2(z)
\end{eqnarray}
where $K_2(z)$ is the second kind Bessel function. 
$Y_{leq}$ describes lepton abundance at thermal equilibrium,
\begin{eqnarray}
Y_{leq}= \frac{6}{s_N}\frac{m_{N}^3\xi(3)}{4\pi^2}
\end{eqnarray}
where $\xi(x)$ is the Riemann zeta function. According to our model, dark matter production is mainly due to the process $NN\rightarrow \phi_{dm}\phi_{dm}$ during leptogenesis non-thermally, and $Y_{dm} \approx 0$ at early Universe. The cross-section of $f\bar{f}\rightarrow Z_p\rightarrow NN$ is too small and makes little contribution to leptogenesis in the model, where $f$ is the fermion of the model. The term $\Delta W $ on the right side of the Boltzmann equation represents the possible scattering processes that wash out the generated $(B-L)$ asymmetry. These scattering washouts include $lW^{\pm}(Z)\rightarrow NH$, $lZ_p\rightarrow NH$, $lH \rightarrow NW^{\pm}(Z)$, $lH\rightarrow l^cH^*$ and $lN\rightarrow Z_p H$ according to Ref.~\cite{Borah:2021mri}, where $l$ represents lepton in the model and $H$ corresponds to SM Higgs. In addition, the annihilation of $NN \to Z_pZ_p$ can also dilute the baryon asymmetry although the process is  highly suppressed due to the tiny charge $q$. More details about these washout terms can be found in Ref.~\cite{Borah:2021mri}. In this work, such scattering processes can make little difference to leptogenesis due to the tiny $U(1)_{L_{\mu}-L_{\tau}}$ charge.
   
After solving the Boltzmann equations, the lepton asymmetry is converted to the baryon asymmetry through sphaleron transitions with the conversion rate \cite{Khlebnikov:1988sr}
\begin{eqnarray}
&&Y_B=-\frac{8N_f+4N_H}{22N_f+13N_H}
Y_{B-L}=-\frac{8}{23}Y_{B-L}
\end{eqnarray}
where we have $N_f=3$ as the number of fermion families and $N_H=2$ as the number of Higgs doublets in our model.

\begin{figure}[h]
\centering
\includegraphics[width=7cm,height=5cm]{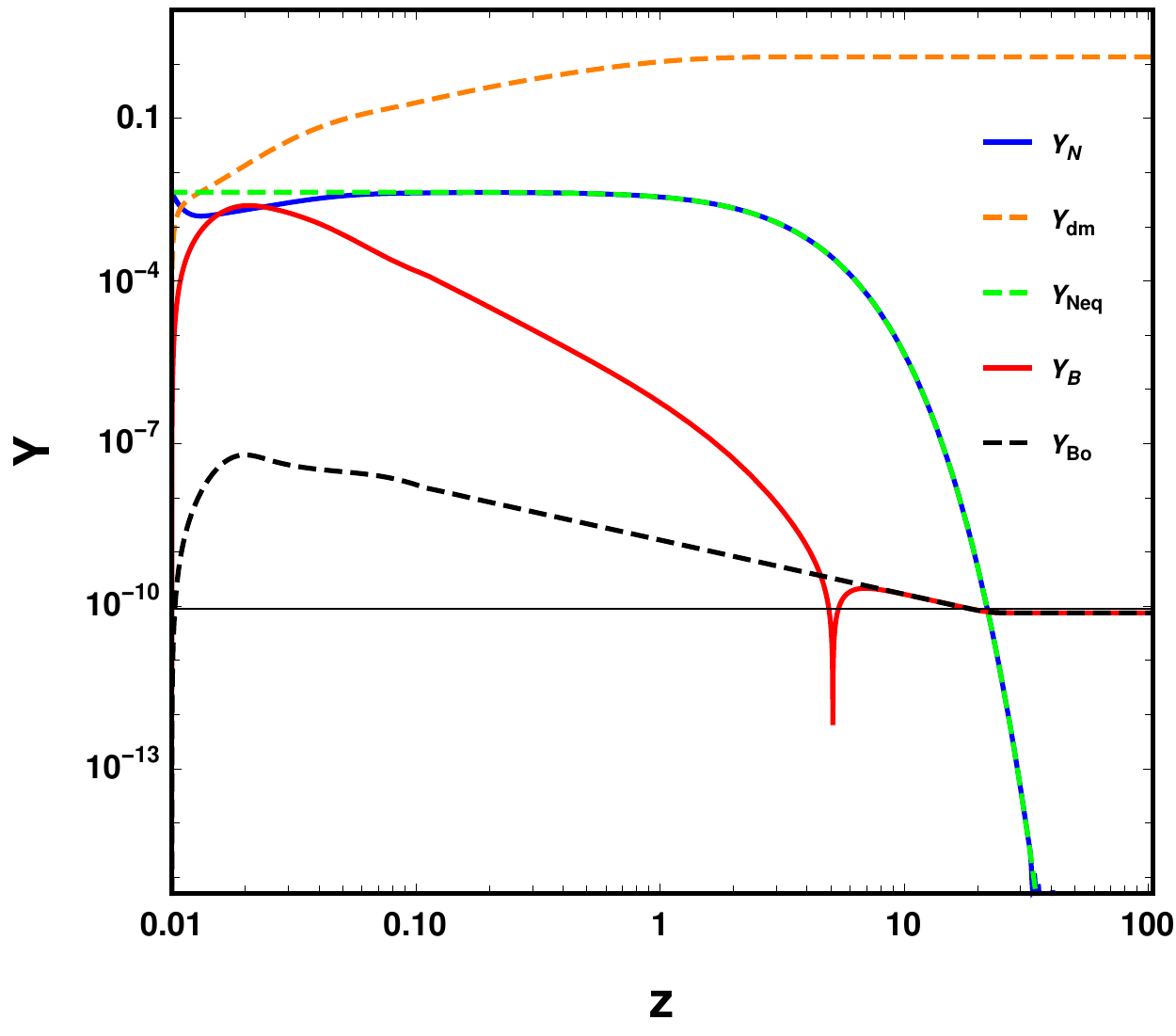}
\includegraphics[width=7cm,height=5cm]{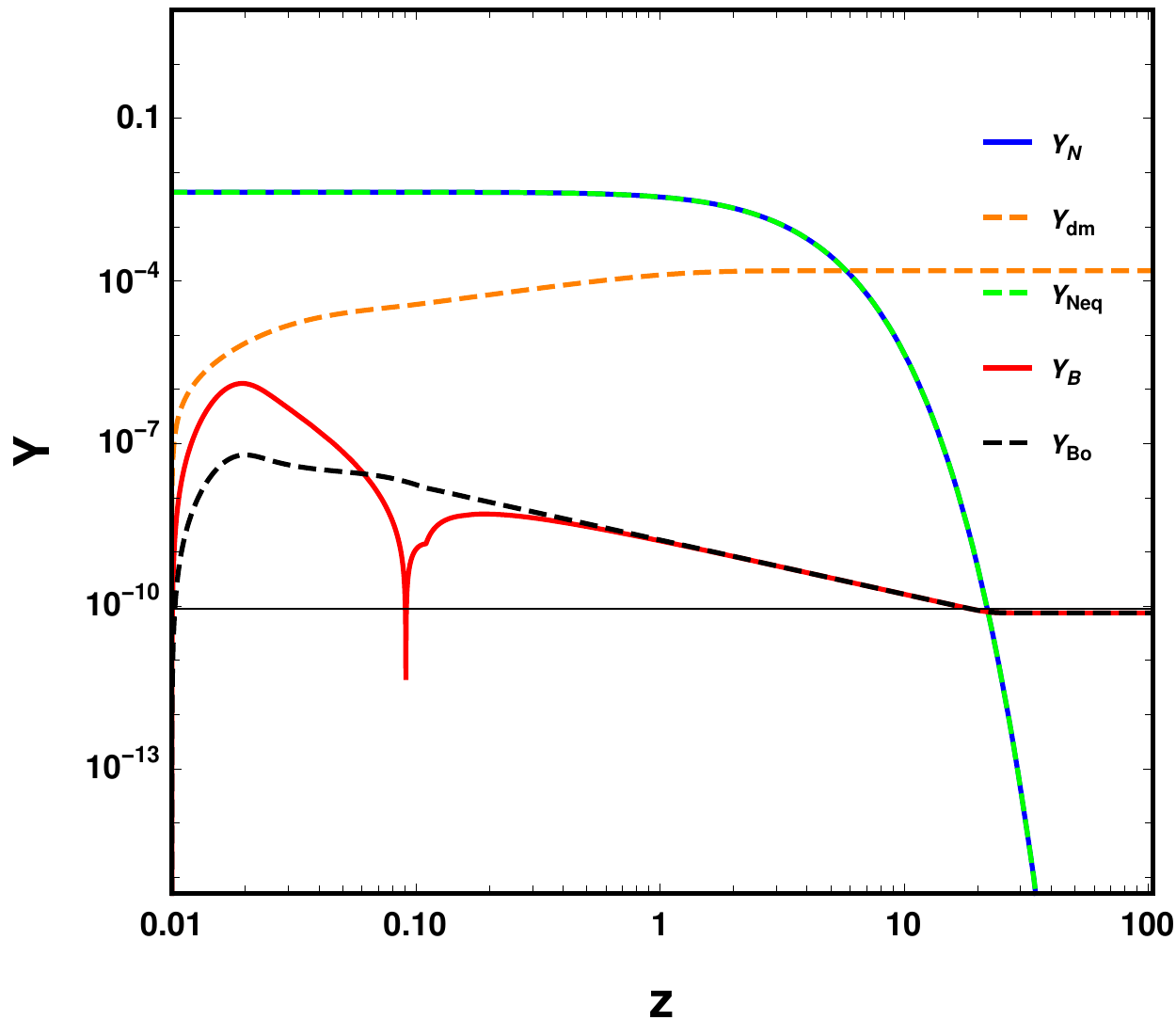}
\includegraphics[width=7cm,height=5cm]{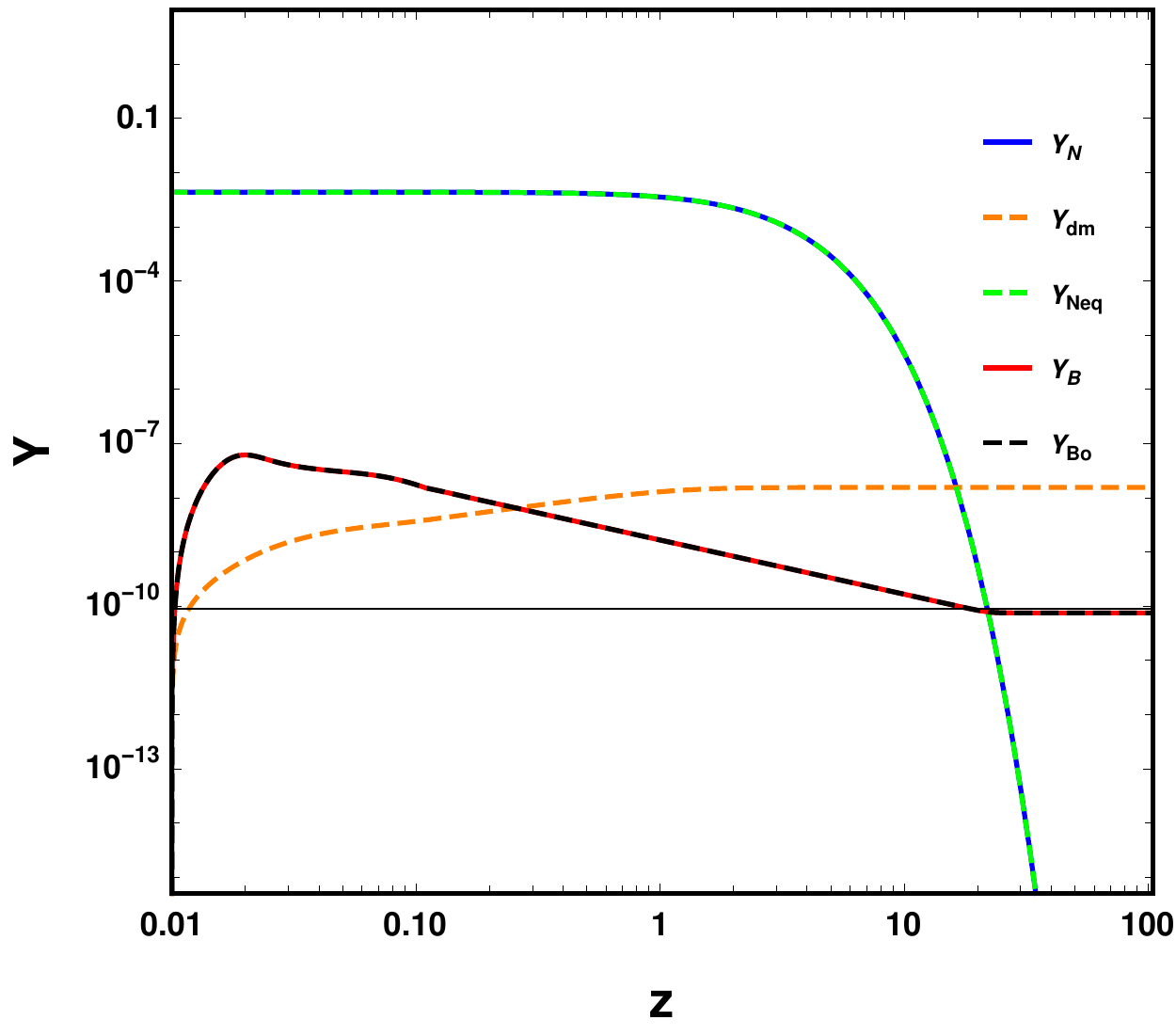}
\caption{
Evolution plots of $N$ abundance $Y_{N}$, dark matter abundance $Y_{dm}$ and Baryon asymmetry $Y_B$, where the black dashed lines $Y_{Bo}$ in all three pictures are the results without dark matter and the grey line corresponds to the observed value according to Planck 2018 \cite{Planck:2018vyg}. 
In the top picture we fix $y_{st}=0.5$ and $m_{\phi_{dm}}=10^{-4}$ GeV, 
in the second picture we set $y_{st}=0.05$ and $m_{\phi_{dm}}=10^{-4}$ GeV, and in the bottom picture we set $y_{st}=0.005$ and $m_{\phi_{dm}}=10^{-2}$ GeV.}
\label{Fig:fig3}
\end{figure}

The evolution of Boltzmann equations are given in Fig.~\ref{Fig:fig3}, where  the grey lines correspond to the observed value according to Planck 2018 \cite{Planck:2018vyg} and the black dashed lines $Y_{Bo}$ in the pictures are the results without dark matter, which are obtained by solving Boltzmann equations of right-handed neutrino. In the first picture we fix $y_{st}=0.5$ and $m_{\phi_{dm}}=10^{-4}$ GeV, in the second picture we set  $y_{st}=0.05$ and $m_{\phi_{dm}}=10^{-4}$ GeV, and in the third one we set $y_{st}=0.005$ and $m_{\phi_{dm}}=10^{-2}$ GeV.

According to the reaction rate of $NN\rightarrow \phi_{dm}\phi_{dm}$, dark matter abundance is related to the coupling $y_{st}$ as well as dark matter mass $m_{\phi_{dm}}$. In the  first and second pictures of Fig.~\ref{Fig:fig3}, we have much exceeded $Y_B$  density at $z \approx 0$ compared with the case without dark matter. The reason for the extra $Y_B$ is that too many dark matter particles have been generated so the inverse process is possible. It is obvious that we also have a peak in the first two pictures. The peak in the two pictures arising from the large reaction rate at $z \approx 0.03$ region due to the large $y_{st}$, such process dilutes the $N$ production so that decreases the $B-L$ asymmetry. For smaller $y_{st}$, the contribution of the process related to dark matter is much small compared with the $N$ decay and the baryon asymmetry almost coincides with the case without dark matter. The orange lines in Fig.~\ref{Fig:fig3} are the evolution of dark matter abundance in our model and for large Yukawa coupling $y_{st}$, dark matter abundance increases a lot due to the large reaction rate. According to Fig.~\ref{Fig:fig3}, the Yukawa coupling $y_{st}$ plays an important role in determining dark matter relic density. Moreover, the final baryon symmetries in the three pictures are almost identical to those in the case without dark matter, suggesting that the process of generating dark matter can make little difference on leptogenesis.

\subsection{Dark matter}
\begin{figure}[h]
  \centering
  \includegraphics[height=5cm,width=5cm]{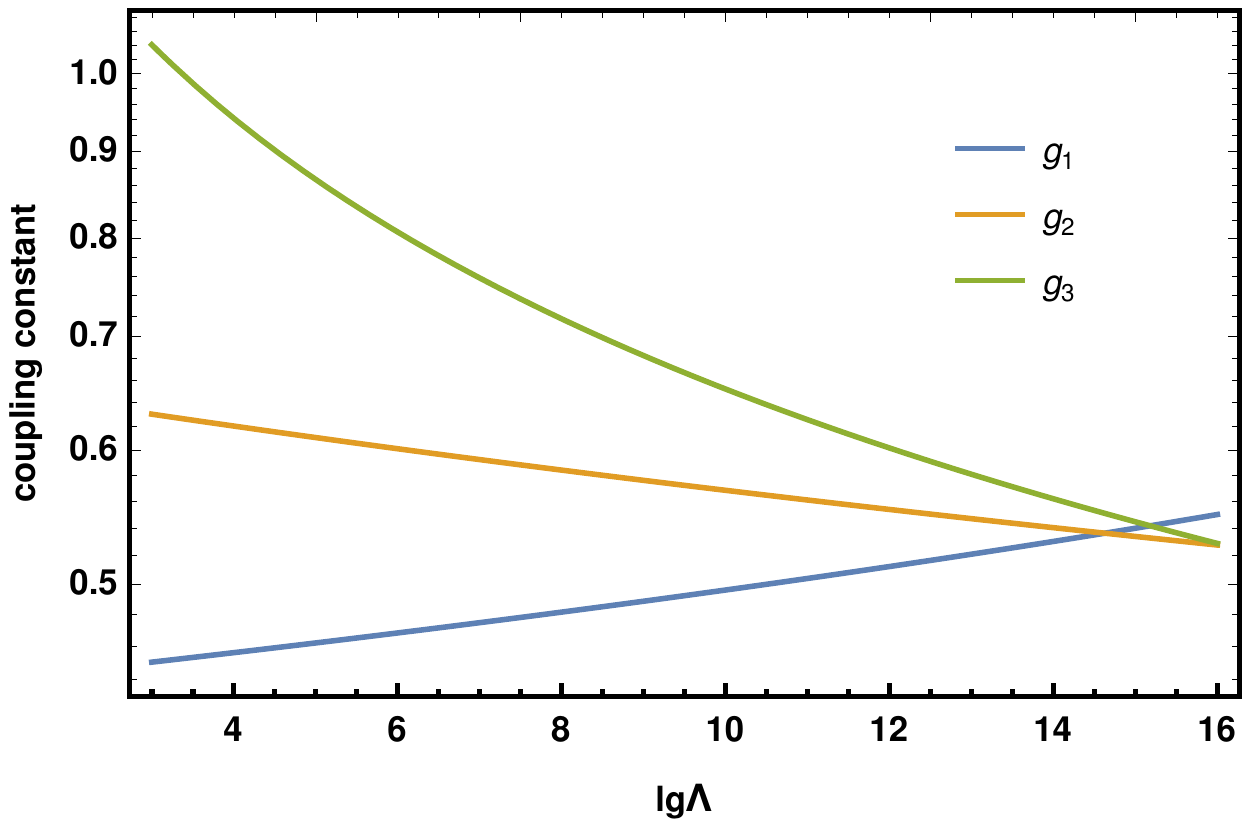}
  \caption{Renormalisation group evolution of $g_1$,$g_2$ and $g_3$,where we fix $g_p=0.01$,$y_{st}=0.01$, $g_1=0.45$,$g_2=0.63$ and $g_3=1.04$ for the initial value, the x-axis correspond to $lg \Lambda$ with $\Lambda$ being the energy scale.}
  \label{Fig:3e}
  \end{figure}

Before discussing the dark matter, we identify the justification for our idea. Firstly, given that dark matter is very light while the mother particle RHNs are very heavy, the huge hierarchy in scales of dark matter production and low energy phenomenology requires appropriate renormalization  group evolution (RGE) in order to make sense of low energy observables. In fact, we obtain the RGEs with the help of SARAH \cite{Staub:2013tta}, and in Fig.~\ref{Fig:3e} we give the RGEs of $g_1$,$g_2$ and $g_3$ corresponding to the coupling constants of $U(1)_Y$, $SU(2)_W$ and $SU(3)_c$,  where we fix $g_p=0.01$,$y_{st}=0.01$, $g_1=0.45$,$g_2=0.63$ and $g_3=1.04$ for the initial value, and the x-axis corresponds to $lg\Lambda$ with $\Lambda$ being the energy scale. The results are consistent with the SM case.
  
  Then, we turn to the non-thermal mechanism to generate dark matter relic density. Firstly, the number density of $N$ is Boltzmann suppressed when the temperature $T< m_N$, which means dark matter is  mainly generated in the early universe with $T >m_N$ for the heavy right-handed neutrino mass. Roughly speaking, one can estimate the rate comparison for the process $NN \to \phi_{dm}\phi_{dm}$ and the expansion rate of the universe $H(T)$ \cite{Bento:2001yk} with,
    \begin{eqnarray}
    H(T)=(\frac{4\pi^3g_*}{45})^{1/2}\frac{T^2}{m_{pl}}=1.66\times g_*^{1/2}\frac{T^2}{m_{pl}}
    \end{eqnarray}
 while the reaction rate $\Gamma$ can given by
\begin{eqnarray}
 \Gamma =n<\sigma v> \sim y_{st}^4T
\end{eqnarray}
 where we have used the temperature as the center-of-mass energy. The   temperature in thermal equilibrium can be given by:
 \begin{eqnarray}
  T{eq} \sim \frac{y_{st}^4m_{pl}}{1.66g_*^{1/2}}
\end{eqnarray}
 Therefore, dark matter will not be thermalized as long as $T_{eq}<m_N$ and we have
\begin{eqnarray}
y_{st} < \sqrt[4]{\frac{m_N}{m_{pl}}}
\end{eqnarray}
 For $m_N>10^{10}$ GeV, one can estimate $y_{st}$ is approximately smaller than $\mathcal{O}(0.01)$. 

As we have  discussed above, dark matter abundance in our model is generated with the process $NN\rightarrow \phi_{dm}\phi_{dm}$ during leptogenesis. Current experiment analysis gives dark matter density with \cite{Planck:2018vyg},
   \begin{eqnarray}
    \Omega h^2 =0.120\pm 0.001
   \end{eqnarray}
and it gives a strict constraint on the parameter space of the dark matter model. The dark matter density expression can be estimated as followed \cite{Mohapatra:2020bze},
\begin{eqnarray}
 \Omega h^2 \approx \frac{m_{\phi_{dm}}s_0Y_{dm}(\infty)}{\rho_0 /h^2}
\label{eq36}
\end{eqnarray}
where $\rm s_0=2890/cm^3$ is the entropy density of the present universe, $\rho_c/h^2=1.05 \times 10^{-5}$ GeV$\rm cm^3$ is the critical density, and $Y_{dm}(\infty) \approx Y_{dm}(m_N/T_{sph})$ with $T_{sph}$ being sphaleron temperature.
It is worth stressing that the dominant constraint on the dark matter arising from its free streaming length, which describes the average distance when a particle travels without collisions, and the expression of free streaming length can be given by \cite{Falkowski:2017uya},
\begin{eqnarray}
  r_{FS}=\int_{a_{rh}}^{a_{eq}} \frac{\langle v\rangle}{a^2H(a)}da \approx \frac{a_{NR}}{H_0\sqrt{\Omega_R}}(0.62+\log(\frac{a_{eq}}{a_{NR}})) \ \ 
\end{eqnarray}
where $H(a)$ is the Hubble parameter, $a(t)$ is the FRW scale factor, $\langle v\rangle$ is the averaged velocity of dark matter $\phi_{dm}$, $a_{eq}$ and $a_{rh}$ correspond to scale factors in equilibrium and reheating respectively, and $a_{NR}$ is the non-relativistic scale factor. According to \cite{Planck:2015fie}, we can apply the result of $ H_0=$ $\rm 67.3\ kms^{-1} Mpc^{-1}$, $\Omega_R=9.3 \times 10^{-5}$ and $a_{eq}=2.9 \times 10^{-4}$. Therefore, the non-relativistic scale factor $a_{NR}$ for the scalar dark matter $\phi_{dm}$ in our model can be given by \cite{Planck:2015fie},
\begin{eqnarray}
 a_{NR}=\frac{T_0}{m_{\phi_{dm}}}(\frac{g_{*,0}}{g_{*,{rh}}})^{1/3}K^{-1/2}
\end{eqnarray}
where $K$ is the decay parameter defined by \begin{eqnarray}
 K=\tilde{m}/m_*
\end{eqnarray}
where $\tilde{m}$ and $m_*$ is the effective neutrino mass \cite{Plumacher:1996kc}, defined as 
\begin{eqnarray}
 \tilde{m}=\frac{(y^{\dagger}y)_{33}}{m_N}
\end{eqnarray}
and $m_*$ is the equilibrium neutrino mass,
\begin{eqnarray}
 m_* \approx 1.08 \times 10^{-3} eV .
\end{eqnarray}
  
We can fix $g_{*,0}=3.91$, $g_{*,{rh}}=106.75$ and $T_0=2.35 \times 10^{-4}$ eV, and we can get \cite{Liu:2020mxj},
\begin{eqnarray} \nonumber
 r_{FS} &\approx& 5.6 \times 10 ^{-2} (\frac{\rm keV}{m_{\phi_{dm}}})(50/K)^{1/2}\times (1+ 0.09 \log[(\frac{m_{\phi_{dm}}}{\rm 2keV})(\frac{K}{50})^{1/2}]) \rm Mpc .
\end{eqnarray}
The most strict constraint on $r_{FS}$ results from small structure formation $r_{FS}$  $<0.1$ Mpc  \cite{Berlin:2017ftj}. 
     The case of  $\ r_{FS}>0.1$ Mpc corresponds to hot DM scenario, while  $ 0.01$  Mpc $<r_{FS}<0.1$  Mpc and $ r_{FS}<0.01$ Mpc correspond to warm as well as cold  DM scenario  \cite{Merle:2013wta}.

\subsection{Combined results}

In this part, we scan a viable parameter space to discuss the interplay between dark matter and leptogenesis.  We choose  $y_{De}$, $y_{D\mu}$, $y_{D\tau}$, $v_0$, $y_{st}$ and $m_{\phi_{dm}}$ as inputs. For simplicity, we set $y_{De}= 0.06$ and we scan a viable parameter space with,
\begin{eqnarray} \nonumber
&&  y_{D\mu},y_{D\tau} \subseteq[10^{-4},1], \\ \nonumber
&&  v_0 \subseteq[1\rm \ GeV, 246\rm \ GeV), \\ \nonumber
&&  m_{\phi_{dm}} \subseteq[10^{-6} \rm \ GeV, 0.1\rm \ GeV], \\
&&  y_{st} \subseteq[10^{-5},0.5] .
\end{eqnarray}
   The right-handed neutrino masses are determined by $y_{De,D\mu,D\tau}$ and $v_0$, and we bring these parameters in the expression of light neutrino mass and reverse solving $M_R$ matrix so that the eigenvalues of $M_R$ are approximately equal to the right-handed neutrino masses. We restrict the parameter space to satisfy baryon asymmetry and dark matter relic density constraint, where the final baryon asymmetry $Y_{B}$ is within $3\sigma$ range of the observed value with $Y_{B} \in [8.60,8.84]\times 10^{-11}$ and $0.11 <\Omega  h^2 <0.13$, and the results are given as followed.

\begin{figure}[h]
\centering
\includegraphics[width=8cm,height=5cm]{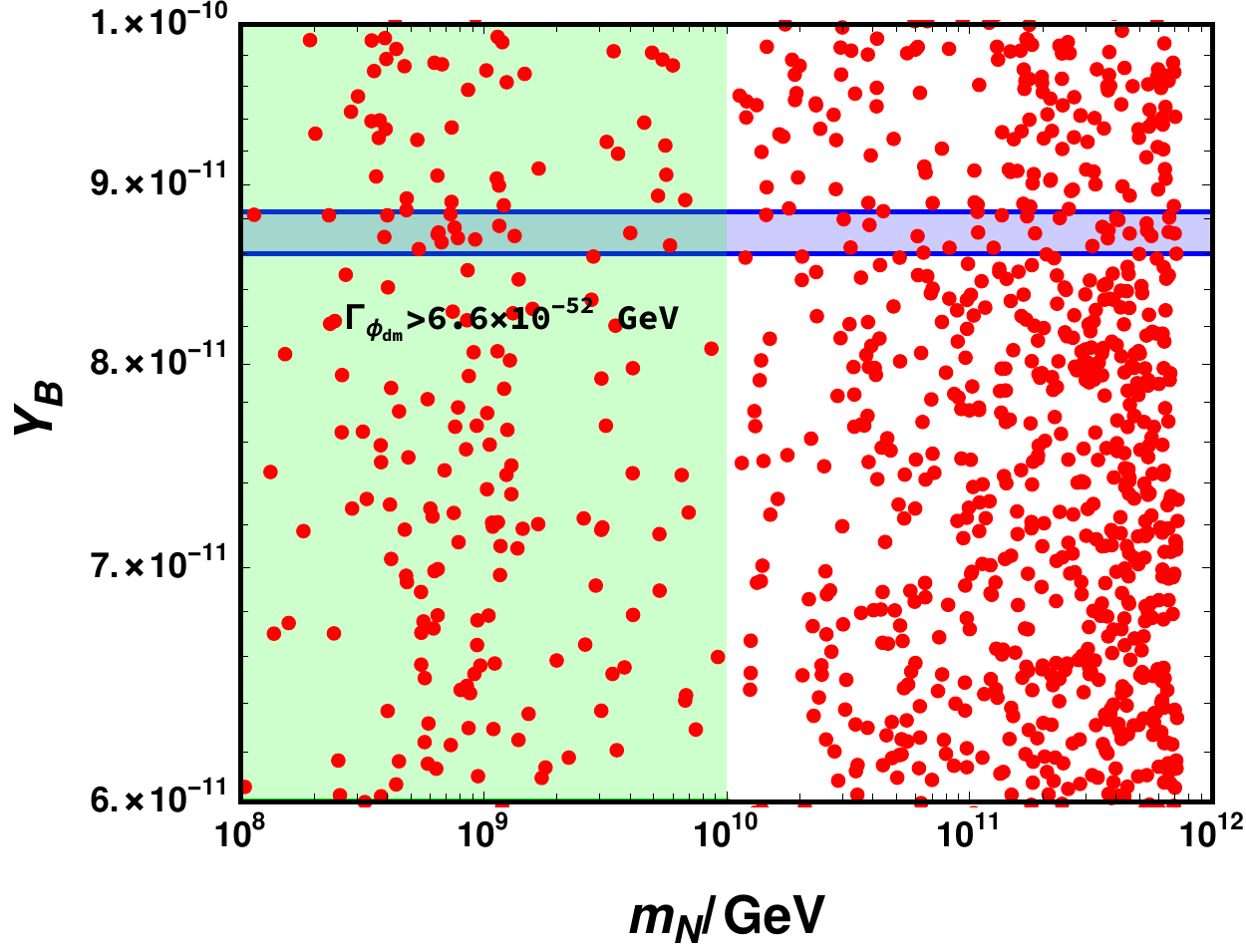}
\caption{Result of $m_N-Y_{B}$ within the chosen parameter space, where the blue band corresponds to the final baryon asymmetry  within $3\sigma$ range of the observed value with $Y_{\Delta B} \in [8.60,8.84]$. The light-green region is excluded due to the dark matter stability condition.}
\label{Fig:fig4}
\end{figure}

\begin{figure}[h]
\centering
\includegraphics[width=7cm,height=5cm]{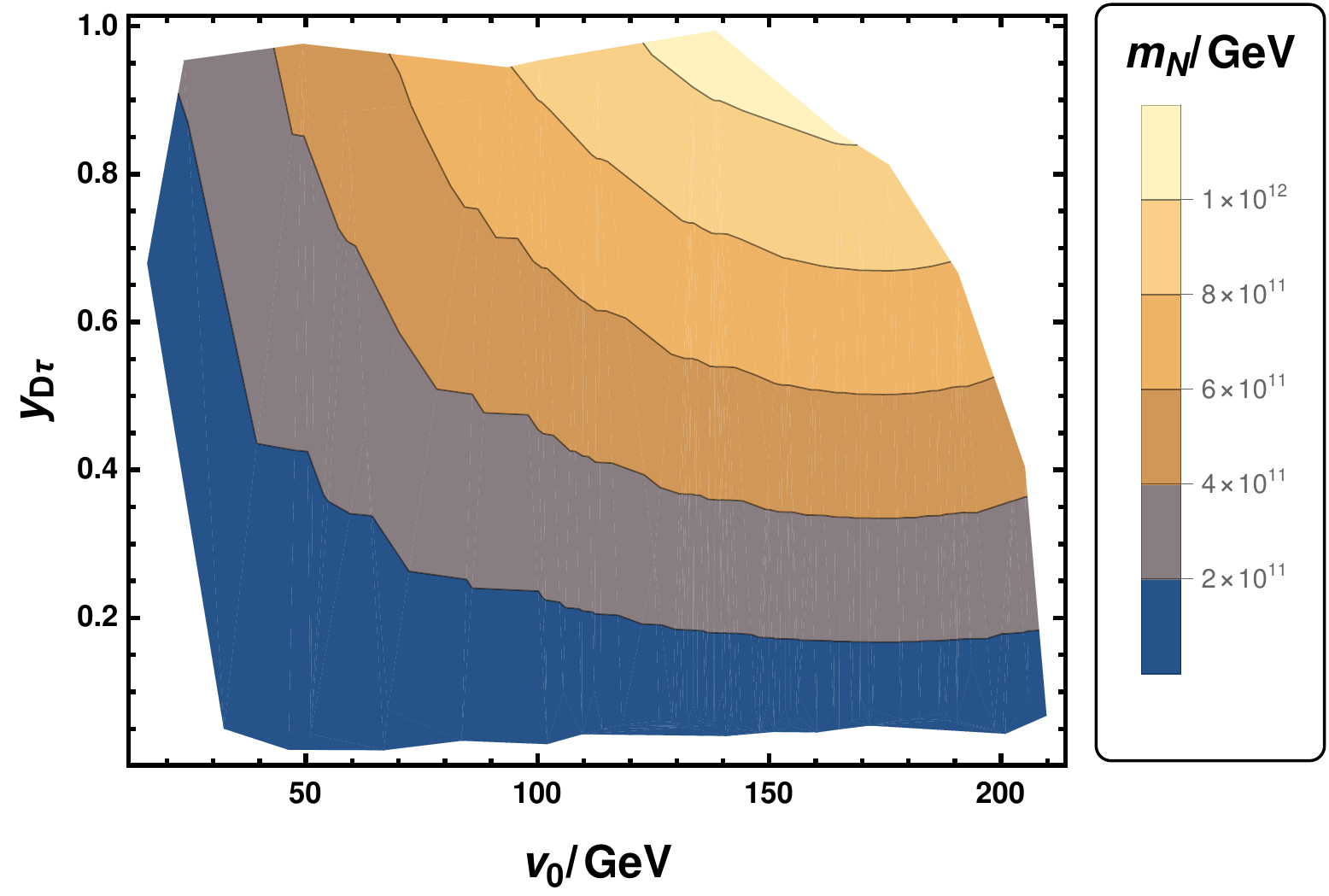}
\caption{Contour plot of $v_0-y_{D\tau}$ satisfying baryon asymmetry as well as dark matter relic density constraint, where the x-axis corresponds to $v_0$ and the y-axis corresponds to $y_{D\tau}$.}
\label{Fig:fig5}
\end{figure}

In Fig.~\ref{Fig:fig4}, we give the result of $m_N-Y_{B}$ within the chosen parameter space, where the blue band corresponds to the final baryon asymmetry  within $3\sigma$ range of the observed value, and the light-green region with $m_N<10^{10}$ GeV is excluded due to dark matter stability constraint. According to Fig.~\ref{Fig:fig4}, right-handed neutrino mass is limited within $\rm [10^8\ GeV,10^{12}\ GeV]$, and one can always have a successful leptogenesis within the chosen parameter space. However, dark matter can decay into a pair of neutrinos due to the heavy-light neutrino mixing in our model, and a stable dark matter demands $m_N>10^{10}$ GeV to ensure the mixing angle small enough so that $\Gamma_{\phi_{dm}}< 6.6\times 10^{-52}$ GeV in the case of $y_{st} \subseteq[10^{-5},0.5]$.
Furthermore, in Fig.~\ref{Fig:fig5}, we give the contour plot of $v_0-y_{D\tau}$ satisfying baryon asymmetry and dark matter relic density constraint with $m_N >10^{10}$ GeV.  According to Fig.~\ref{Fig:fig5},  for $v_0< 15$ GeV, the corresponding right-handed neutrino mass is lighter than $10^{10}$ GeV regardless of the value of $y_{D\tau}$. Similarly, for $y_{D\tau}<0.02$, right-handed neutrino mass is always lighter than $10^{10}$ GeV regardless of the value of $v_0$. Therefore, a viable parameter space satisfying leptogenesis as well as dark matter stability constraint is $v_0\geq 15$ GeV and $y_{D\tau}\geq 0.02$ for $m_N >10^{10}$ GeV.

\begin{figure}[h]
\centering
\includegraphics[width=7.5cm,height=5cm]{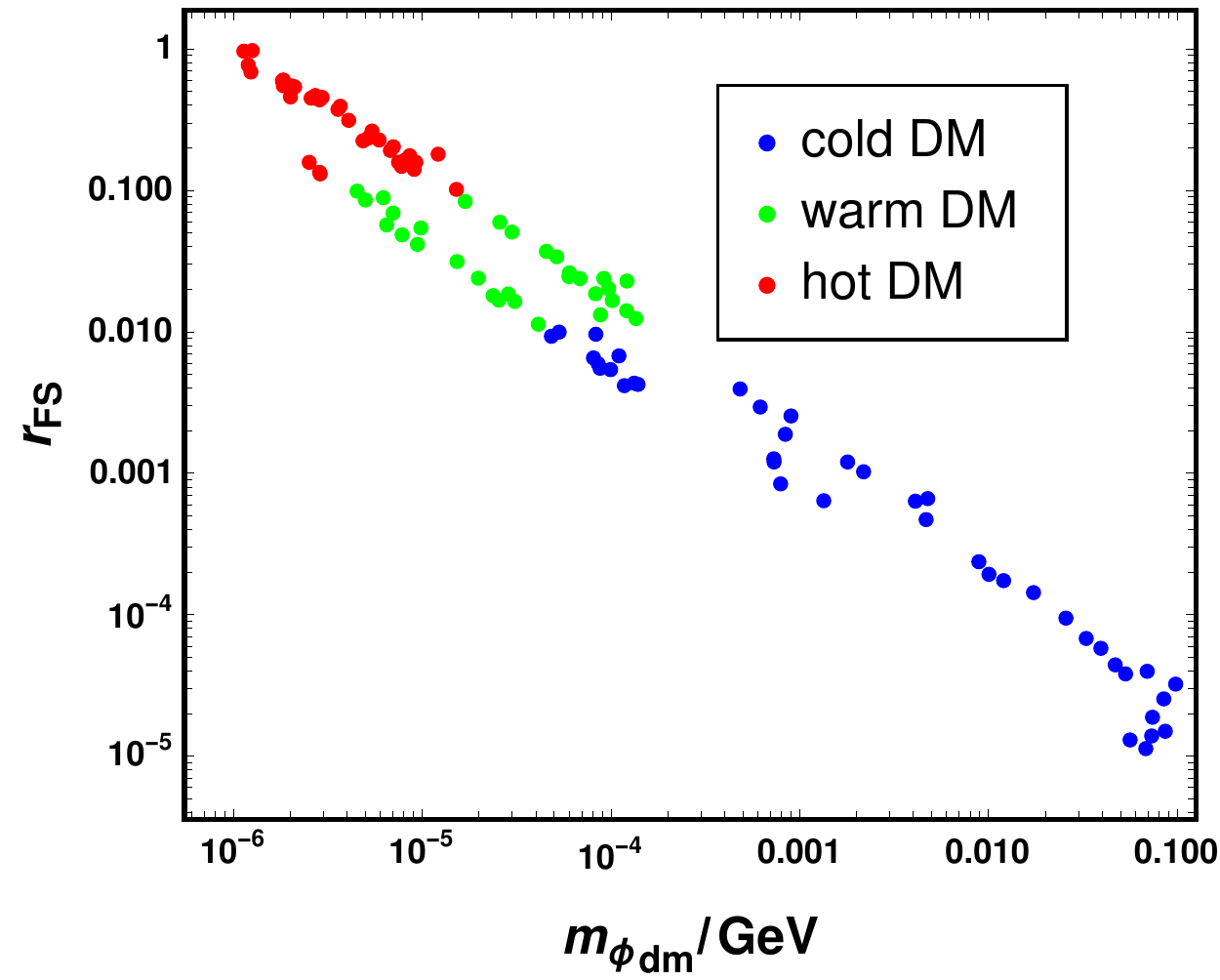}
\caption{Parameter space of $m_{\phi_{dm}}-r_{FS}$ satisfying baryon asymmetry and dark matter relic density constraint. The blue points correspond to cold dark matter case with $ r_{FS}<0.01$ Mpc, the green points correspond to warm dark matter case with $ 0.01$ Mpc $< r_{FS}< 0.1$ Mpc, and the red points correspond to hot dark matter case with $ r_{FS} > 0.1$ Mpc.}
\label{Fig:fig6}
\end{figure}

 In addition, We can split the viable parameter space into three scenarios as in Fig.~\ref{Fig:fig6}, according to the constraint from free streaming length $r_{FS}$. The hot DM scenario is excluded by small structure formation. For warm DM, the possible dark matter mass region is $ 10^{-5}$ GeV $<m_{\phi_{dm}} < 10^{-4}$ GeV. Meanwhile for cold DM, $ 10^{-4}$ GeV $<m_{\phi_{dm}} < 0.1$ GeV is allowed. Moreover, $r_{FS}$ decreases to about $10^{-5}$ Mpc in the case of $m_{\phi_{dm}} \sim 0.1$ GeV.

\begin{figure}[h]
\centering
\centering
\includegraphics[width=7.5cm,height=5cm]{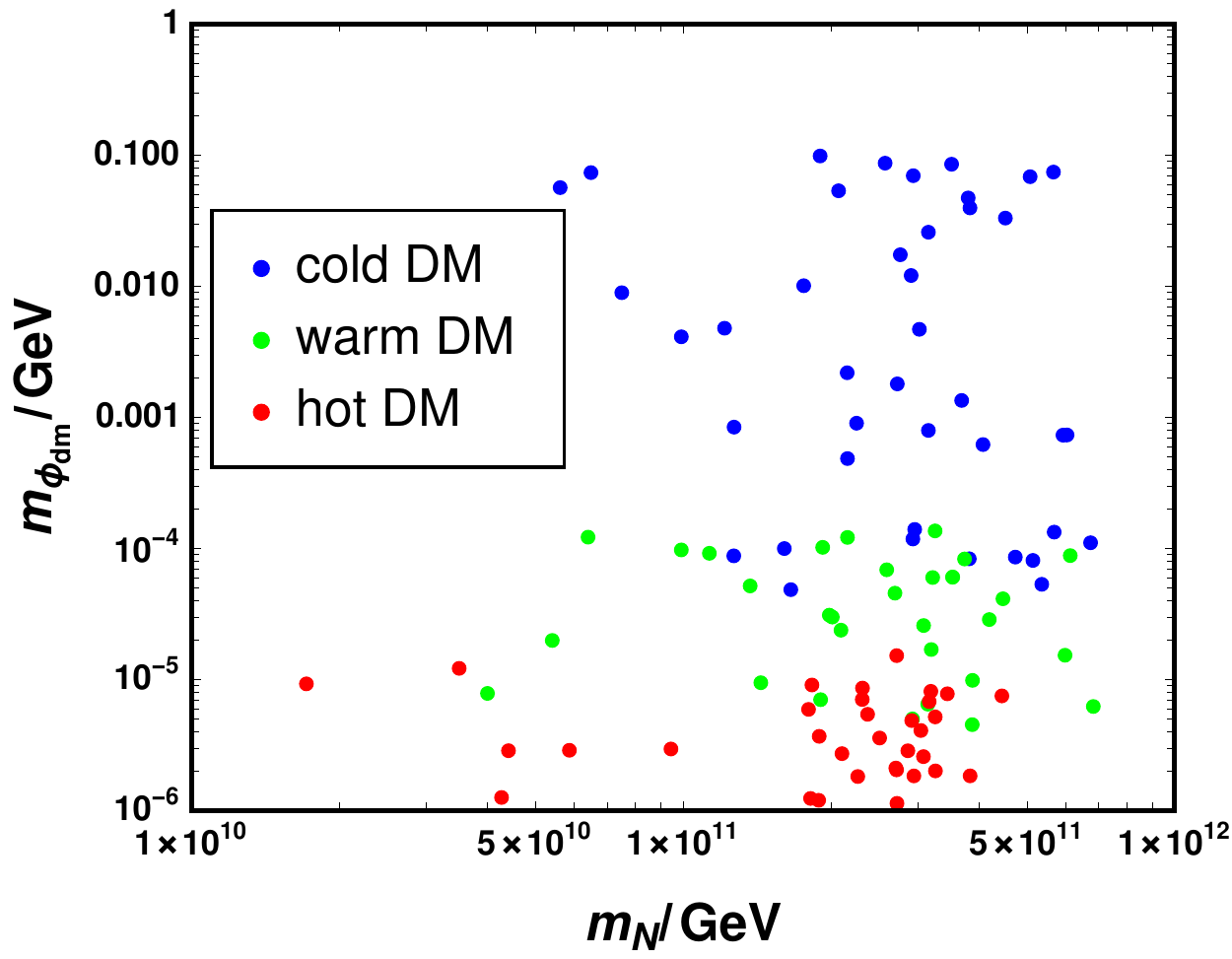}
\caption{Parameter space of $m_N-m_{\phi_{dm}}$ satisfying baryon asymmetry and dark matter relic density constraint. The blue points correspond to cold dark matter case with $ r_{FS}<0.01$ Mpc, the green points correspond to warm dark matter case with $ 0.01$ Mpc $< r_{FS}< 0.1$ Mpc, and the red points correspond to hot dark matter case with $ r_{FS} > 0.1$ Mpc. }
\label{Fig:fig7}
\end{figure}

\begin{figure}[h]
\centering
\includegraphics[width=7.5cm,height=5.3cm]{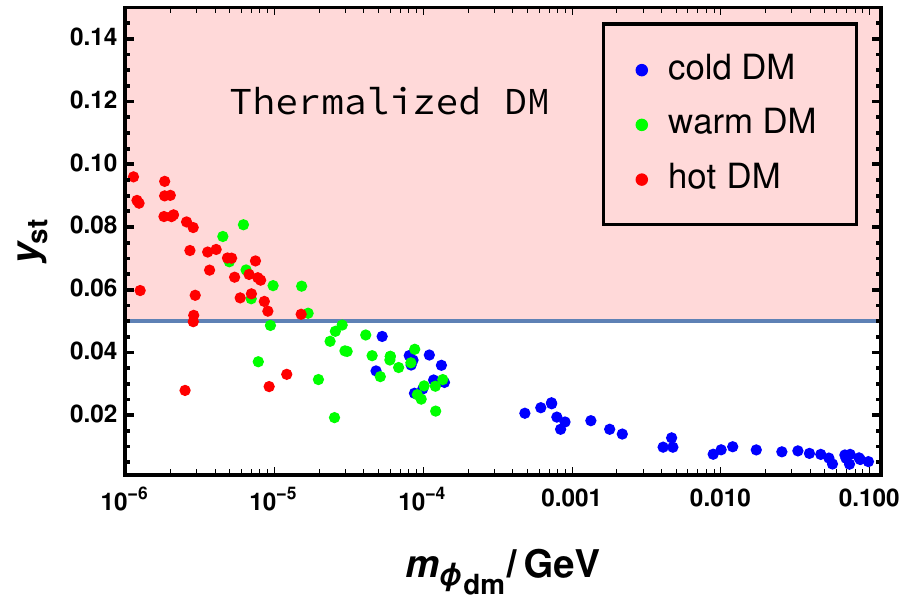}
\caption{Parameter space for dark matter satisfying baryon asymmetry and  relic density constraint. The blue points correspond to cold dark matter case with $ r_{FS}<0.01$ Mpc, the green points correspond to warm dark matter case with $ 0.01$ Mpc $< r_{FS}< 0.1$ Mpc, and the red points correspond to hot dark matter case with $ r_{Fs} > 0.1$ Mpc.}
\label{Fig:fig8}
\end{figure}
We show the parameter space of $m_N-m_{\phi_{dm}}$ in Fig.~\ref{Fig:fig7}, where the colored points represent parameter space  that satisfies the baryon asymmetry constraint  as well as dark matter relic density constraint. The heavy right-handed neutrino mass is constrained within $ \rm (1 \times 10^{10}\ GeV,1 \times 10^{12}\ GeV) $ while the dark matter mass is $\rm (10^{-6}\ GeV, 0.1\ GeV)$. Moreover, according to Fig.~\ref{Fig:fig6}, we have different colored points corresponding to hot, warm and cold dark matter separately, and three kinds of dark matter are all possible for a certain value of $m_N$ such as $m_N \sim 5\times10^{11}$ GeV. Meanwhile, it is interesting that a  heavier $m_N$  often indicates warm or cold DM production. In fact, this is because dark matter is mainly generated during leptogenesis, and heavier $m_N$ means a higher temperature and leptogenesis starts earlier so that dark matter has a long sufficient cooling time.
 
Dark matter abundance in our model depends on the singlet Yukawa interactions and the dark matter parameter space of $m_{\phi_{dm}}-y_{st}$ is important in our model which can determine the correct relic density. According to Eq.~\ref{eq25} and Eq.~\ref{eq36}, dark matter abundance is approximately proportional to $y_{st}^4m_{\phi_{dm}}$, which means $m_{\phi_{dm}}$ and $y_{st}$ are stringently restricted by the relic density constraint. In Fig.~\ref{Fig:fig8}, we give the result of $m_{\phi_{dm}}-y_{st}$ and the colored points correspond to hot, warm and cold DM scenarios respectively. We find a low bound for $y_{st}$ with $ y_{st}\geq 0.004$. For smaller $y_{st}$, dark matter production is not enough compared with the observed value. Moreover, a lighter dark matter always demands a larger $y_{st}$ so that a larger cross-section is to obtain the correct relic density.
It is worth emphasizing  that for $y_{st}>0.05$ the cross-section is so large that dark matter will be thermalized in the late time, which corresponds to the light-red region in Fig.~\ref{Fig:fig8}. Therefore, a viable parameter space for $y_{st}$ is $0.004<y_{st}<0.05$  for the non-thermal production. For the general two-sector leptogenesis scenario, sterile neutrino being dark matter can be generated by the decay of right-handed neutrino  via freeze-in mechanism and dark matter abundance depends on the branching ratio of right-handed \cite{Falkowski:2017uya}, which corresponds to a wide parameter space. However, in our work, the Yukawa coupling $y_{st}$ related to scalar dark matter production is constrained strictly within a narrow region although dark matter mass $10^{-5}$ GeV $<m_{\phi_{dm}}<0.1$ GeV is allowed. Last but not least, the contribution of dark matter to the final baryon asymmetry can be negligible due to the small cross-section of $NN \rightarrow \phi_{dm}\phi_{dm}$, and the baryon asymmetry constraint does not exclude any parameter. Therefore, we do not need to discuss it elsewhere. 
\section{summary}
\label{sec:so}

Two-sector leptogenesis scenario offers a shared mechanism to unify baryon asymmetry and dark matter problems within a common framework, which can also explain the origin of the tiny neutrino mass problem. In this work, we consider the scalar dark matter model within an extended $L_{\mu}-L_{\tau}$ model. Such a model involves neutrino mass, dark matter and can also explain the muon $(g-2)$ anomaly problem. We discuss the non-thermal dark matter production generated by right-handed neutrinos during the leptogenesis process.  The main sector that restricts the stability of dark matter is that dark matter can decay into a pair of neutrinos due to the heavy-light neutrino mixing, which requires the right-handed neutrino $N_{\tau}$ mass larger than $10^{10}$ GeV. We scan a viable parameter space that satisfies baryon asymmetry and dark matter relic density constraint, where  $N_{\tau}$ mass  is constrained within the region of $\rm (1 \times 10^{10}\ GeV, 1 \times 10^{12}\ GeV)$ consistent with the leptogenesis.  On the other hand, $N_{\tau}$ mass is also related to the decay parameter $K$, which can determine whether dark matter is hot, warm or cold DM and the hot dark matter case is not favored by small structure formation. We found the allowed dark matter mass region is $\rm (10^{-5}\ GeV, 0.1\ GeV)$. Moreover, Yukawa coupling $y_{st}$ related to dark matter can not be necessarily too small and the  region to generate the right dark matter relic density is $0.004<y_{st}<0.1$. However, for too large  $y_{st}$, one can have the thermalized dark matter case, therefore the right region for $y_{st}$ is $(0.004,0.05)$.

\section*{Acknowledgements}
We thank Wei Liu for the early collaboration and useful
communications on this work. Hao Sun is supported by the National Natural Science Foundation of China (Grant No.12075043, No.12147205).

\bibliographystyle{JHEP}
\bibliography{qxnew}
\end{document}